\documentclass{aa}  
\usepackage{graphicx}
\usepackage{txfonts}
\begin{document}

   \title{A novel approach for the direct estimation of the instantaneous Earth rotation velocity}

   \author{O. Titov
                    }

   \institute{Geoscience Australia,
              PO Box 378, Canberra, ACT, 2601, Australia\\
              \email{oleg.titov@ga.gov.au}          
             }

   \date{Received January .., 2025}

  \abstract
{\emph{Context}.Very Long Baseline Interferometry (VLBI) measures two standard observables: group delay and fringe frequency (delay rate). While group delay is widely used to estimate a 
broad range of geodetic and astrometric parameters, fringe frequency has, to date, been largely ignored.
Here, we show that the fringe frequency is a unique tool for direct estimation of the instantaneous Earth angular rotation velocity, which is not accessible with the group delay alone.\\
\emph{Aims}. We estimate the magnitude of the Earth's angular rotation velocity using a 30-year set of VLBI data and obtain daily estimates of $X$ and $Y$ angles linking the Instantaneous Rotation Pole (IRP) and the International Celestial Reference System (ICRS) pole.\\
\emph{Methods}. The plain least-squares method was applied to analyse the fringe frequency available from routine geodetic VLBI observations. \\
\emph{Results}. We estimate three components of the Earth rotation vector on a daily basis with a formal error of 1 prad/s or $10^{-8}$ in relative units, or better, if a large international VLBI network is at work. The newly obtained values can be used to monitor the Earth's rotation irregularity in parallel to the traditional length-of-day (LOD) values and to directly assess the modern precession-nutation theory.}
 

   \maketitle

\section{Introduction}

Geodetic Very Long Baseline Interferometry (VLBI) measures the difference in arrival times of radio waves ‒ referred to as ‘group delay’ ‒ at two radio telescopes receiving signals from distant sources, and a relative fringe frequency (also known as delay rate) which compensates for the geocentric motion of each radio telescope due to the Earth's rotation. This measurement achieves remarkable precision, with accuracies of 20-40 picoseconds (ps) for the group delay and 1-100 frad for the fringe frequency \citep{Schuh12}.
VLBI observations are 
conducted in two frequency bands: 2.3 GHz (S-band) and 8.4 GHz (X-band), where the lower 
frequency serves to calibrate ionospheric fluctuations present in the X-band data.
By combining data from a network of radio telescopes distributed globally, this technique 
can determine Earth orientation parameters (EOP) with an accuracy of approximately 0.1 
milliarcseconds (mas) during a standard 24-hour geodetic VLBI experiment \citep{Bizouard2019}. The positions of 
the radio telescopes, in Cartesian coordinates, are provided by the International Terrestrial 
Reference Frame (ITRF), while the coordinates of reference radio sources are taken from the 
International Celestial Reference Frame (ICRF3) catalogue.
Historically, the group delay 
measured by geodetic VLBI has been used to derive polar motion relative to the Celestial Intermediate Pole (CIP) within the 
ITRF, utilising three parameters: the two components of polar motion (x- and y-pole motion) 
and Universal Coordinated Time (UTC). 
The CIP is linked to the Geocentric Celestial Reference Frame (GCRF) pole through a precession-nutation model. This traditional concept relies on the position of the equinox, which is steadily moving along the ecliptic due to the motion of the Earth's rotational axis with respect to the distant celestial objects, with a period of about 26,000 years.
The alternative method proposes an introduction of the non-rotating origin (NRO) instead of the equinox point \citep{Guinot1979}. To connect the instantaneous rotation pole and the celestial reference frame, new angles ($X$ and $Y$) must be introduced; these are linked to the precession-nutation parameters \citep{Capitaine1990}. The theory links the position of the so-called celestial intermediate pole (CIP) with the celestial pole offsets $X$ and $Y$, and the corrections ($dX$ and $dY$) should be estimated. \cite{Fukushima2001} showed that the NRO adopted by the International Astronomical Union (IAU) leads to the appearance of a global secular rotation of the celestial ephemeris origin with a rate of -4.15"/year, and suggested additional theoretical options \citep{Fukushima2003}. However, this NRO approach has not been widely used to date due to a lack of observations that could directly provide the $X$ and $Y$ values. Here, we show that VLBI delay rate observations produce three components of Earth's angular rotation vector, which can be used to calculate the $\Omega$ magnitude. The conversion of these components to the celestial pole positions ($X$ and $Y$ angles) will be presented in future publications.

\section{Basic VLBI delay rate equations}

The fringe frequency measured with VLBI reflects 
the generalised Doppler effect, which describes the frequency shift between an ‘emitter’ and a 
‘receiver’ moving relative to each other and to the direction of signal propagation. The resultant 
frequency shift is the difference between the two Doppler shifts observed at each radio telescope. 
Since the signal originates from the same radio source, the contribution from the ‘emitter’ 
cancels out, leaving only the contributions from the two ‘receivers’. It should be noted that the fringe frequency is calculated simultaneously with the group delay during the correlation process, rather than derived from the group delay \citep{Sovers_1998}. The primary term of the 
fringe frequency can be expressed as shown by \cite{Cohen_1971}:

\begin{equation}
	\frac{\Delta f_{21}}{f}=-\frac{(w_{2}\cdot s)-(w_{1}\cdot s)}{c},
	\label{Eq1} % Use a logical label
\end{equation}

where $w_{1}$  and  $w_{2}$ are vectors of the geocentric velocities of the radio telescopes with geocentric positions assigned by vectors $r_{1}$  and  $r_{2}$, $s$ is the unit vector in the direction of the radio source, $c$ is the speed of light, and (·) denotes the dot product. This analytical equation Eq.~(\ref{Eq1}) can be considered as the derivative $\frac{\Delta f_{21}}{f}=\frac{d\tau}{dt}$ of the group delay, $\tau=-\frac{(b_{21},s)}{c} = -\frac{(r_{2}\cdot s)-(r_{1}\cdot s)}{c} $, followed by the complete relativistic equation provided in Appendix A (again noting that the observed value $\frac{\Delta f_{21}}{f}$ is not a derivative of the observed group delay).  For rotational motion, the linear velocity vector is calculated as $w =  [\Omega \times r]$, where $\Omega$ is the vector of the Earth's angular rotation ($\Omega = (\omega_{x}, \omega_{y},\omega_{z})$, $|\Omega| \approx 7.2921151467\cdot 10^{-5} rad/s$), $r$ is the vector of the radio telescope geocentric position, and $[\times]$ denotes the vector cross product. Equation (\ref{Eq1}) can then be written as

\begin{equation}
	\frac{\Delta f_{21}}{f}=-\frac{([\Omega\times b_{21}]\cdot s)}{c} = -\frac{([s\times b_{21}]\cdot \Omega)}{c}, 
	\label{Eq2} % Use a logical label
\end{equation}

where the baseline vector $b_{21}=r_{2}-r_{1}$ now appears explicitly. The magnitude of the effect $\frac{\Delta f_{21}}{f}$ for the Earth-based baseline is approximately 0.1 ÷ 1 $\mu$rad. A full relativistic equation for the fringe frequency includes additional terms describing the effects of special and general relativity Eq.(\ref{sup_Eq1}) and Eq.(\ref{sup_Eq3}) to the level of 1 frad  =$10^{-15}$ rad.  The corresponding expected formal error of the vector $\Omega$ components is $10^{-12} -10^{-13}$ rad/s, or $10^{-8} - 10^{-9}$ in relative units, $\frac{\Delta\Omega}{\Omega}$. 

\subsection{Solution setup}

The positions of the celestial intermediate pole in the $X$ and $Y$ directions are predicted by the IAU2000A/IAU2006 nutation and precession models (\citet{Mathews2002}; \citet{Capitaine2003}). The standard approach focusses on analysing group delay and applying corrections to the rotation angle values ($dX$, $dY$) to calculate minor adjustments 
to precession and nutation amplitudes, e.g. (\citet{iers10}; \citet{Belda2017}).  The least-squares method is then employed to solve 
the system of equations after incorporating all relativistic corrections. Typically, this method 
requires a priori information about the estimated EOP. 
While it is often asserted that geodetic techniques cannot measure the $X$ and $Y$ angles directly (\citet{Mendes2009}; \citet{Schreiber2023}), this paper demonstrates that the fringe frequency measured with geodetic VLBI, as described in Eq. (\ref{Eq2}), can effectively fill this gap. 
Eq.~($\ref{Eq2}$) encompasses the complete vector of angular rotation $\Omega$, rather than merely small LOD corrections. 

The Earth's angular rotation vector $\Omega$ defines the positions of the IRP. Thus, the IRP direction could be extracted using the VLBI delay rate and Eq. (\ref{Eq2}). 
However, while the two poles, CIP and IRP, are close enough, their definitions are not equivalent, and the transition of $\Omega = (\omega_{x}, \omega_{y},\omega_{z})$ to the IRP is not discussed in this paper. 

The system of equations for the least-squares method accounts for frequency fringe variations resulting from the non-hydrostatic component of the lower troposphere. The model for 
tropospheric delay rates is detailed in Eq. (\ref{Sup_Eq5}). The positions of reference radio sources, 
geodetic site coordinates, and the EOP were fixed; moreover, a simple parametrical model to adjust the difference in hydrogen maser frequencies, including the constant frequency
offset, was used. Consequently, the model comprises seven parameters per radio telescope (six for the tropospheric delay rate and one for the difference between the two
hydrogen maser frequencies) alongside three components of the vector $\Omega$.

It is important to note that the group delays (used in the standard method) are affected by hydrogen maser phase fluctuation offsets and the local troposphere wet component instability. 
This necessitates the development of complex algorithms to mitigate the effects of phase variations when using standard group delay for data analysis. In contrast, the fringe frequency in Eq. (\ref{Eq2}) is not affected by such phase fluctuations. 
Therefore, a simple least-squares method is effective, yielding daily estimates of the three components of the vector $\Omega$  for each 24-hour VLBI experiment.

We processed VLBI data used in the operational service from April 1993 to April 2024 
(NEOS-A in 1993-2001 and IVS-R1, R4 in 2002-2024) with the OCCAM 
software platform (version 6.3) \citep{Titov2004}
to estimate the three components of the vector $\Omega$ using the aforementioned procedure. All standard reductions of geodetic VLBI data adhere to the IERS Conventions 2010 \citep{iers10}. Station positions and 
velocities were fixed based on ITRF2020 values \citep{Altamimi2023}, with a priori EOP sourced from IERS C 
04 data and the IAU 2000A nutation model. Radio source positions were anchored by the ICRF3 catalogue coordinates \citep{Charlot_2020}. A priori zenith delays were determined from local troposphere parameter values, subsequently mapped to observation elevation using the Vienna mapping function (VMF1) \citep{Boehm2006}.

There are two options for estimating the vector $\Omega$: either add the major contribution (Eq.~\ref{Eq1}) to the full relativistic model, or omit it. In the former case, only small corrections to the three components of the vector $\Omega$ are obtained; in the latter scenario, all three components are estimated in full. In this paper, the latter option is used.

\begin{figure}[ht]
    \centering
    \includegraphics[width=1.0\linewidth]{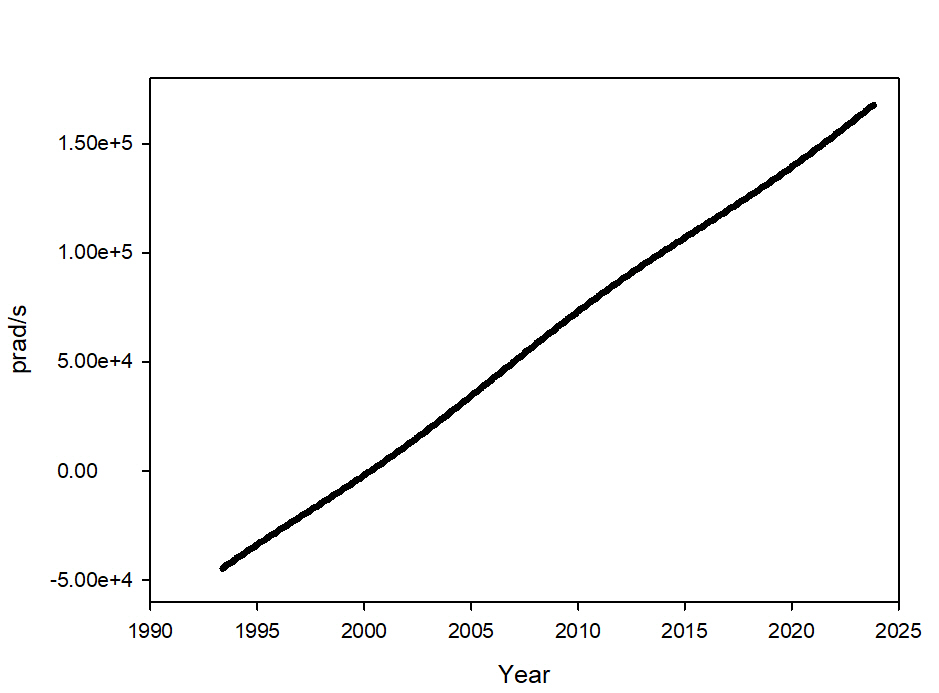}
    \caption{Daily estimates of component $\omega_{X}$ obtained by least-squares method using Eq.~(\ref{Eq2}). }
    \label{omega_x}
\end{figure}

\begin{figure}[ht]
    \centering
    \includegraphics[width=1.0\linewidth]{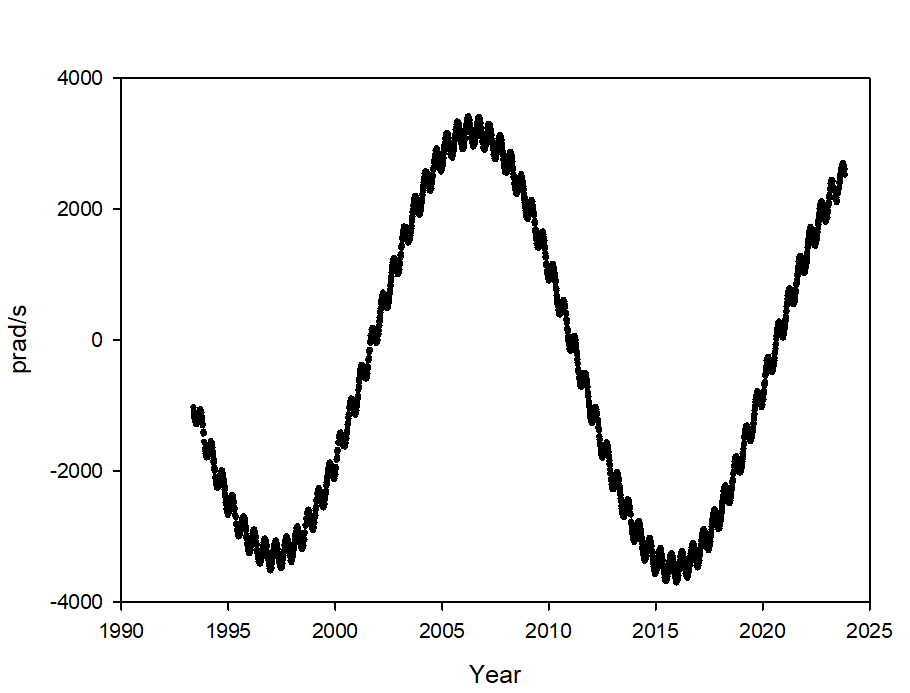}
    \caption{Daily estimates of component $\omega_{Y}$  obtained by least-squares method using Eq.~(\ref{Eq2}).}
    \label{omega_y}
\end{figure}

\begin{figure}[ht]
    \centering
    \includegraphics[width=1.0\linewidth]{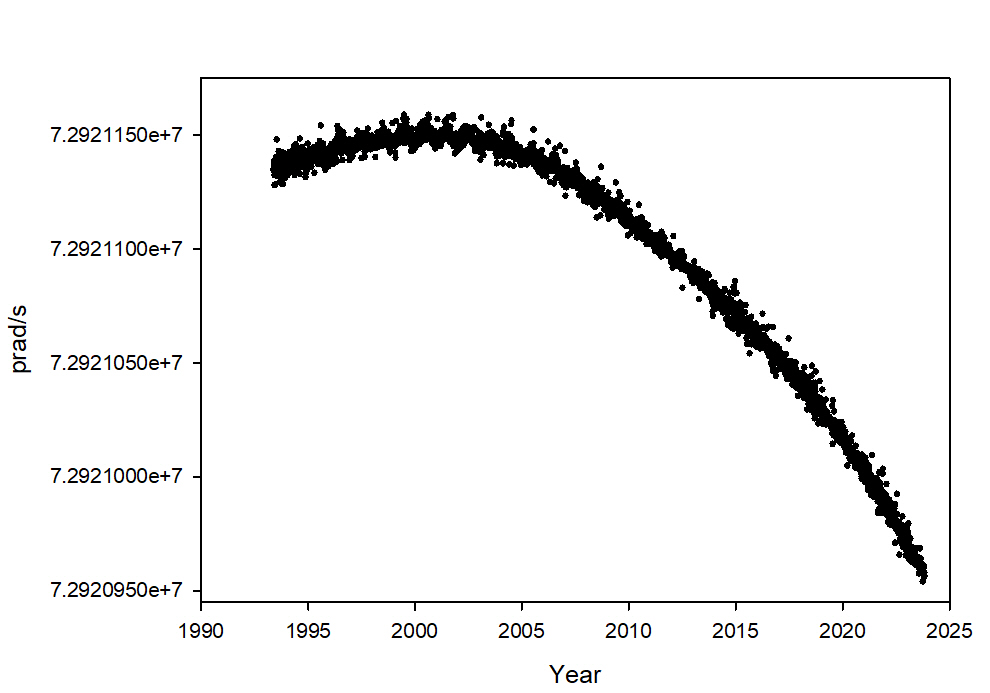}
    \caption{Daily estimates of component $\omega_{Z}$  obtained by least-squares method using Eq.~(\ref{Eq2}).}
    \label{omega_z}
\end{figure}

\begin{figure}[ht]
    \centering
    \includegraphics[width=1.0\linewidth]{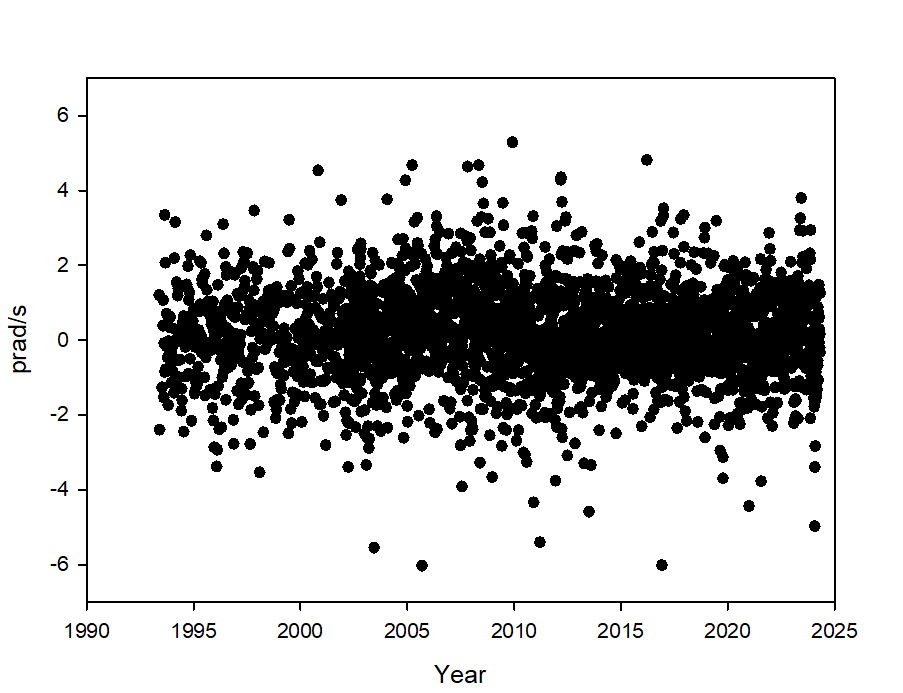}
    \caption{Difference in $\omega_{X}$ between the IERS time series and the new estimates shown in Fig.~\ref{omega_x}.}
    \label{diff_omega_x}
\end{figure}

\begin{figure}[ht]
    \centering
    \includegraphics[width=1.0\linewidth]{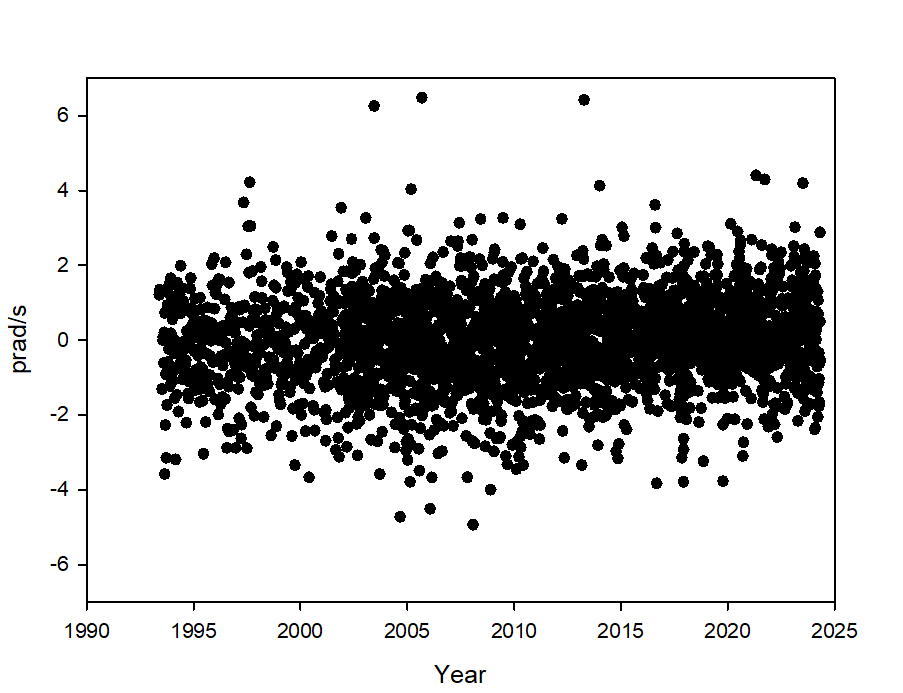}
    \caption{Difference in $\omega_{Y}$ between the IERS time series and the new estimates shown in Fig.~\ref{omega_y}.}
    \label{diff_omega_y}
\end{figure}

\begin{figure}[ht]
    \centering
    \includegraphics[width=1.0\linewidth]{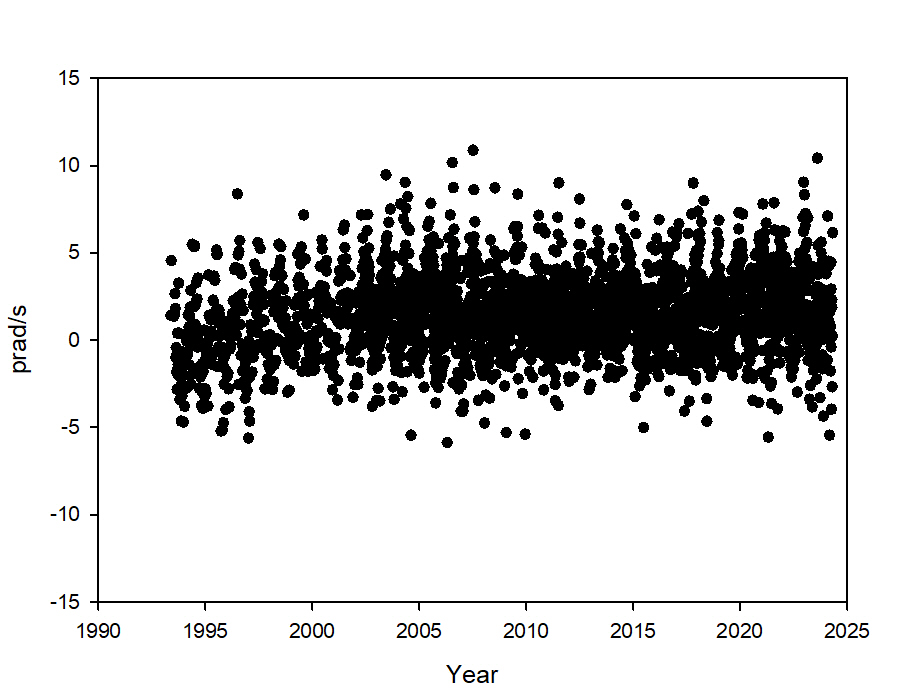}
    \caption{Difference in $\omega_{Z}$ between the IERS time series and the new estimates shown in Fig.~\ref{omega_z}.}
    \label{diff_omega_z}
\end{figure}

\begin{figure}[ht]
    \centering
    \includegraphics[width=1.0\linewidth]{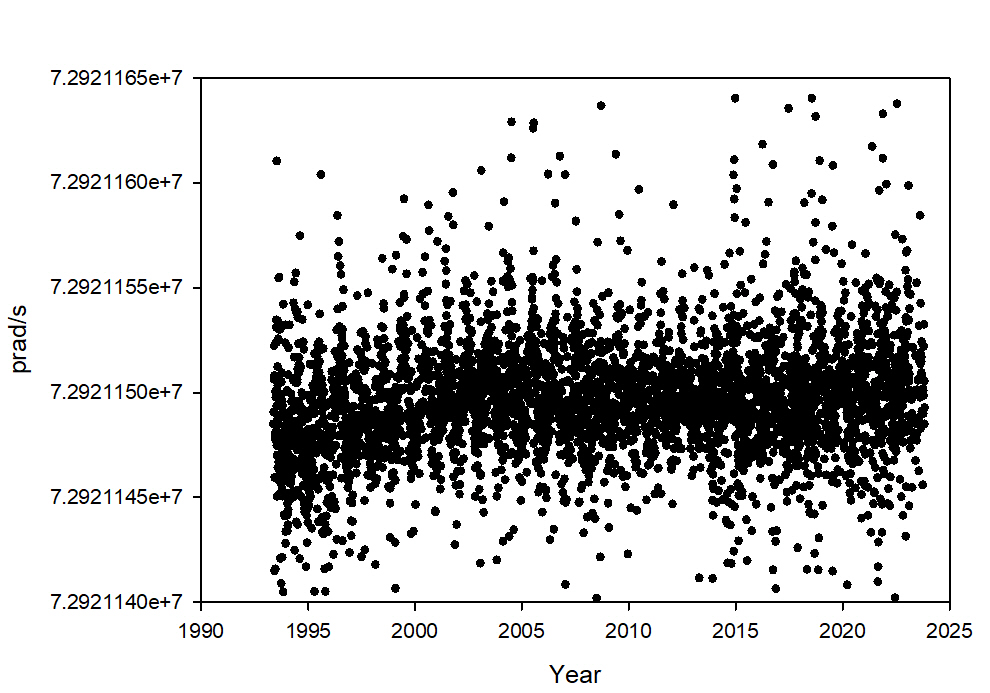}
    \caption{Daily values of the full angular velocity  $\Omega$, derived using Eq.(\ref{Eq3}).}
    \label{omega}
\end{figure}

\section{Results}

Three components of the vector $\Omega = (\omega_{x}, \omega_{y},\omega_{z})$, obtained over the 30-year period of observations, are shown in Figs. \ref{omega_x}, ~\ref{omega_y} and ~\ref{omega_z}, respectively. They represent approximately $1/1000^{th}$ of the magnitude of the 26,000-year circle induced by the precession motion of the rotation pole, overlapped with the main 18.6-year nutation and the seasonal signal. 
The formal errors on all three graphs are approximately 1-10 prad/sec and are too small to be visible in the figures. These results are in good agreement with the time series of the vector $\Omega$
components published by the IERS 
(\footnote{\label{fnote1}\url{https://hpiers.obspm.fr}}). 
~Figs.~4-6 display the differences between the IERS time series and the newly calculated estimates shown in~Figs.~1-3. The root mean square (rms) of the residuals in Figs.~4-6 are 1.2 prad/s, 1.2 prad/s, and 2.7 prad/s for the $X$, $Y$, and $Z$ components, respectively. The results presented in Figs.~1-3 will be discussed in the following three subsections.

\subsection{Effect of general relativity}

\begin{figure}[ht]
    \centering
    \includegraphics[width=1.0\linewidth]{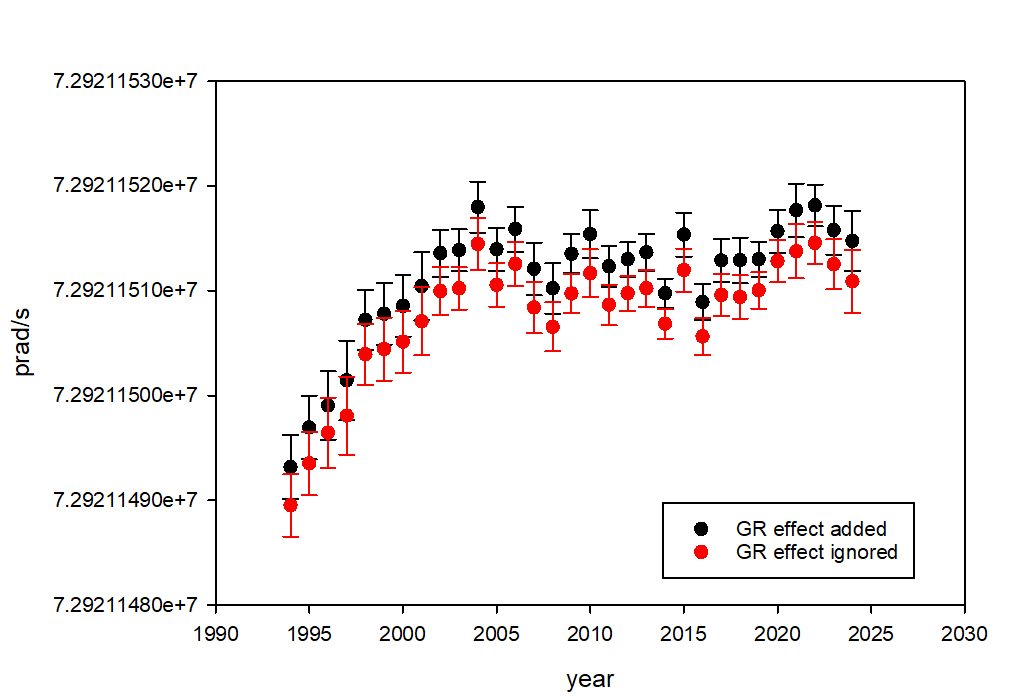}
    \caption{Annual amplitude of the Earth's angular rotation from solutions with and without the contribution of general relativity (black dots and red dots, respectively).}
    \label{fig:GR0}
\end{figure}

\begin{figure}[ht]
    \centering
    \includegraphics[width=0.8\linewidth]{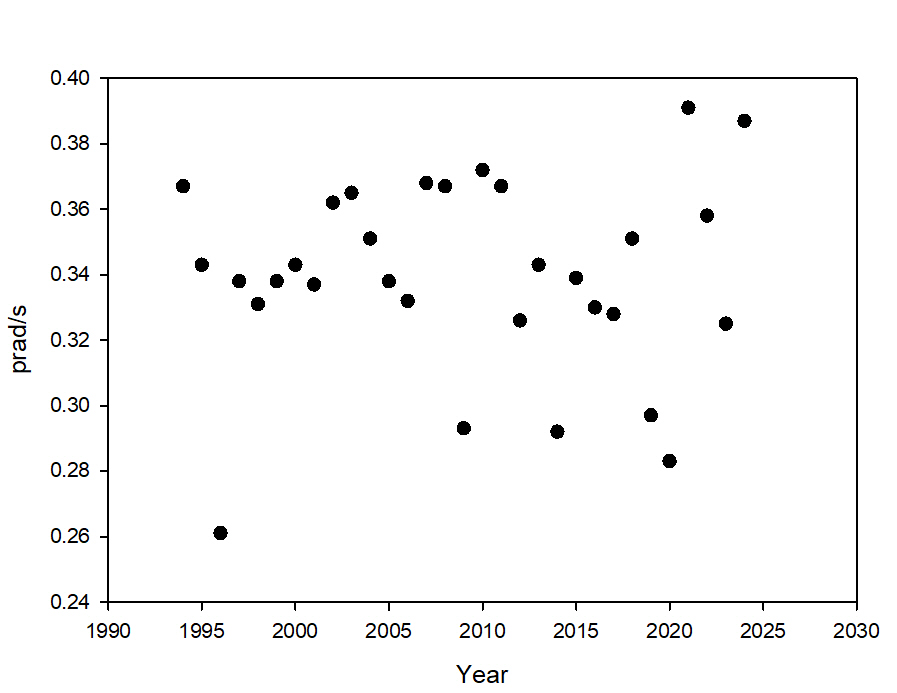}
    \caption{Difference between the two time series shown in Fig.~\ref{fig:GR0}. The
    mean value is  approximately 0.33 prad/s.}
    \label{fig:GR}
\end{figure}

Since the general relativity effect shifts the radio source positions, the analytical model in Eq.~(\ref{sup_Eq3}) should be applied for the fringe frequency reduction. Because the source positions are not estimated within the solution setup, the general relativity effect propagates into other parameters, particularly the Earth's angular rotation velocity.
This effect can optionally be switched off to compare the resultant effect of general relativity on the $\Omega$ and LOD time series. 
Figure~\ref{fig:GR0} presents two sets of time series of the annual angular velocity values, with and without the reduction for the Sun's gravitational field. Figure~\ref{fig:GR} shows the difference between the two graphs in Fig.~\ref{fig:GR0}. The offset varies from 0.25 to 0.40 prad/s for all years between 1993 and 2024, with a mean value approximately 0.33 prad/s. The corresponding offset in LOD is approximately 0.3 msec.
Thus, neglecting the effects of general relativity will lead to an underestimation of the Earth's angular velocity.

\subsection{Angular velocity and length of day}

The magnitude of the angular velocity $\Omega = |\Omega|$ is calculated from Eq.~(\ref{Eq3}) for each 24-hour geodetic VLBI experiment:

\begin{equation}
\Omega = \sqrt{\omega^{2}_{x} + \omega^{2}_{y} + \omega^{2}_{z}}.
	\label{Eq3} % Use a logical label   
\end{equation}

The full magnitude of $\Omega$ is shown in Fig.~\ref{omega} and displays the subtle variations associated with the length of day. Again, the formal errors on the graph are approximately 1-10 prad/sec and not visible here.

To evaluate the quality of the results, the daily values of $\Omega$ can be converted to the official LOD values published by the IERS
(\footnote{\label{fnote1}\url{https://www.iers.org/IERS/EN/DataProducts/EarthOrientationData/eop.html}}).
The equation linking $\Omega$ from Eq. (\ref{Eq3}) and LOD is given, for example, by (\citet{Lambeck}; \citet{Bizouard2022}; \citet{Baenas2021}):

\begin{equation}
\frac{\Delta LOD}{LOD_{0}} = -\frac{\Omega - \Omega_{0}}{\Omega_{0}} ,
	\label{Eq4} % Use a logical label   
\end{equation}

where $\Omega_{0}$ is the so-called ‘nominal’ angular velocity, $\Omega_{0} = 72921151.467$ prad/s, and $LOD_{0}$ = 86400 s.

It is common to link the vector  $\Omega$
components $(\omega_{x}, \omega_{y},\omega_{z})$ to the relative perturbations ($m_{1},m_{2},m_{3}$) induced by external forces (e.g. the Earth's atmosphere) in the form  ($m_{1},m_{2},1 + m_{3}$), so that the angular velocity module is approximated to the first order by  $\Omega(1+m_{3}) $, where $m_{3}$ is the axial change. The relative variations of the length of day in Eq. (\ref{Eq4}) are then  

\begin{equation}
\frac{\Delta LOD}{LOD_{0}} \approx -m_{3}.
	\label{Eq5} % Use a logical label   
\end{equation}

\begin{figure} % Do NOT use \begin{figure*}
	\centering
	\includegraphics[width=0.4\textwidth]{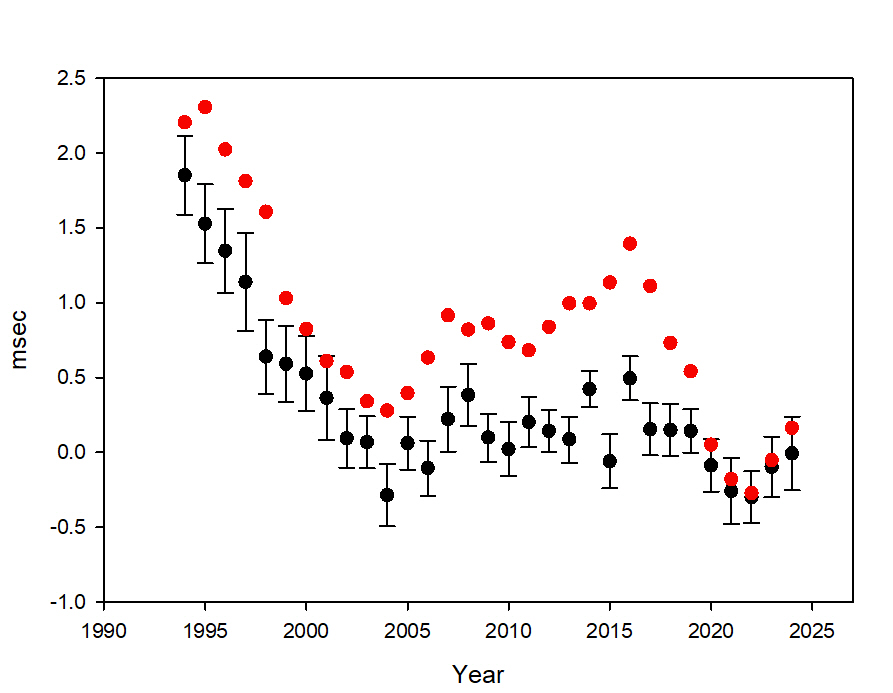} 
    \includegraphics[width=0.4\textwidth]{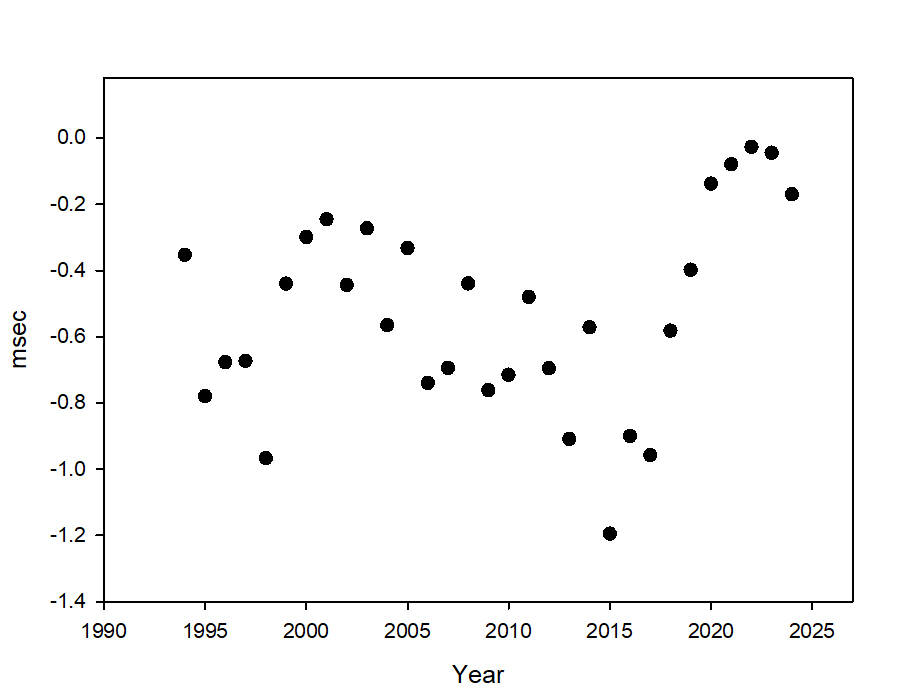}

	% Captions go below figures
	\caption{\textbf{
        Comparison of LOD values.}
Top: yearly time series of LOD values calculated from the angular rotation amplitude ~Eq.(\ref{Eq4}); black dots) and  IERS values (red dots). Bottom: difference between the two LOD time series..}
\label{fig2} % give each figure a logical %label name
\end{figure}

Traditionally, all fluctuations in LOD have been interpreted solely as 
a consequence of axial changes, i.e. $1+m_{3}$. However, a direct comparison of the vector $\Omega$ and LOD, obtained independently, reveals systematic discrepancies that can likely be attributed to the influence of minor constituents typically overlooked in standard LOD approximations~(Fig.~\ref{fig2}). However, a detailed interpretation of these systematics is beyond the scope of this paper.

\subsection{Higher temporal resolution results}

It is often stated in the literature that the geodetic VLBI technique can only achieve $\Omega$ time series with 24-hour resolution (e.g. ~\cite{Schreiber2023}). However, this limitation primarily applies to the standard time series of the EOP produced by the IERS. Much higher temporal resolution for $\Omega$ can be achieved using VLBI fringe frequency data.

\begin{figure} % Do NOT use \begin{figure*}
	\centering
	\includegraphics[width=0.4\textwidth]{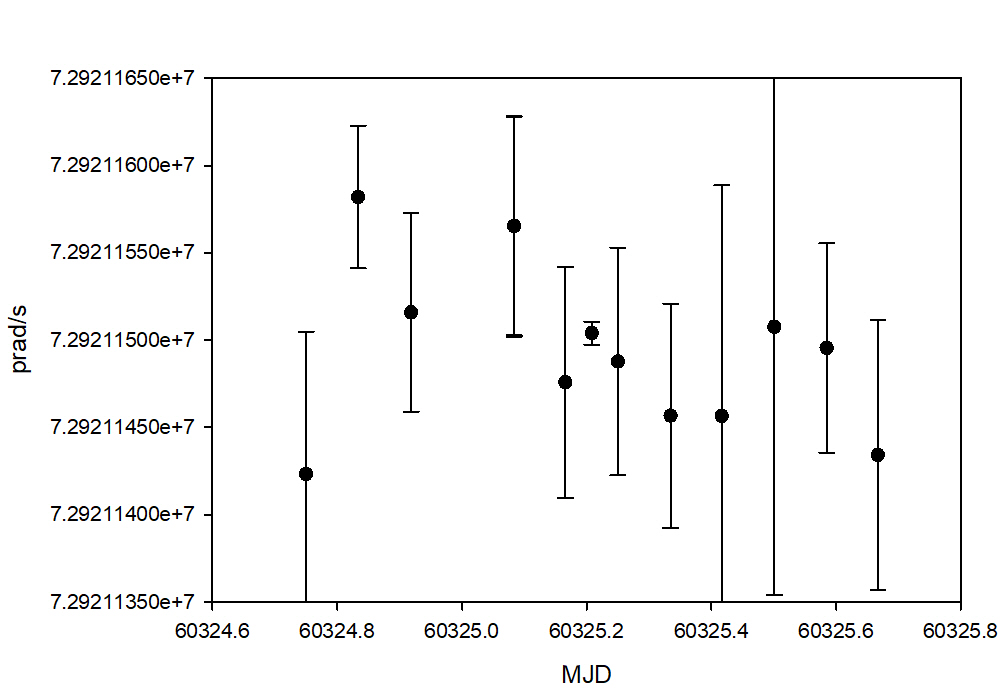} % for an image file named example_figure.*
	    \includegraphics[width=0.4\textwidth]{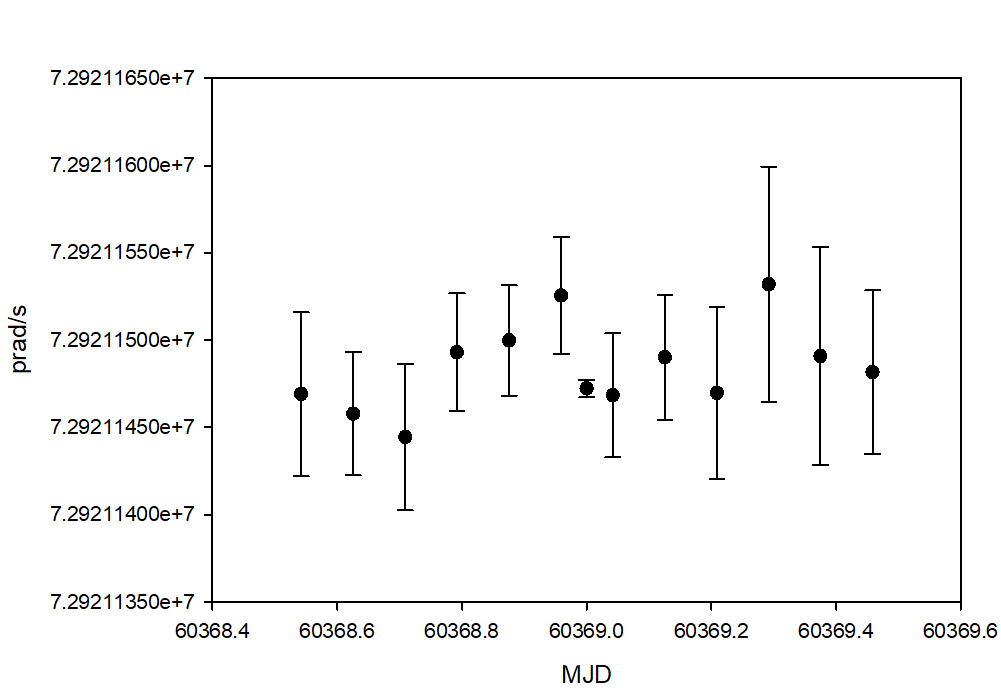}

	% Captions go below figures
	\caption{
        Time series of vector $\Omega$ estimates with 2-hour resolution.
 Top panel: data from the IVS-R1 experiment R1138 on 15 January, 2024.
         Bottom panel: data from the VGOS experiment VO-4059 on 28 February, 2024.
		Daily estimates of the vector $\Omega$ are shown in the middle of each graph and are identified by the smaller formal errors.}
\label{fig3} % give each figure a logical label name
\end{figure}

~Fig.~\ref{fig3} shows the 2-hour time series of the vector $\Omega$ estimates obtained for two recent geodetic VLBI experiments. The first is the standard S/X experiment conducted on 15 January 2024 IVS-R1-1138, and the second is the broadband VGOS experiment on 28 February 2024 (IVS-V0-4059). Since the number of VGOS observations (16,975) is substantially higher than that of S/X (3,940), the average number of observations in each 2-hour interval is around four times larger for the VGOS experiment. The formal errors of the 2-hour estimates vary between 4 and 16 prad/s for the S/X experiment, and between 3 and 7 prad/s for the VGOS experiment, respectively. It should be noted that two hours is not the ultimate limit for the resolution of the $\Omega$ time series. The resolution could be shortened to one hour or even smaller if a dedicated experiment is conducted for the purpose of angular rotation monitoring. Ultimately, this involves a trade-off between time series resolution and $\Omega$ uncertainty for each individual time bin.

\begin{figure}[ht]
    \centering
\includegraphics[width=1.0\linewidth]{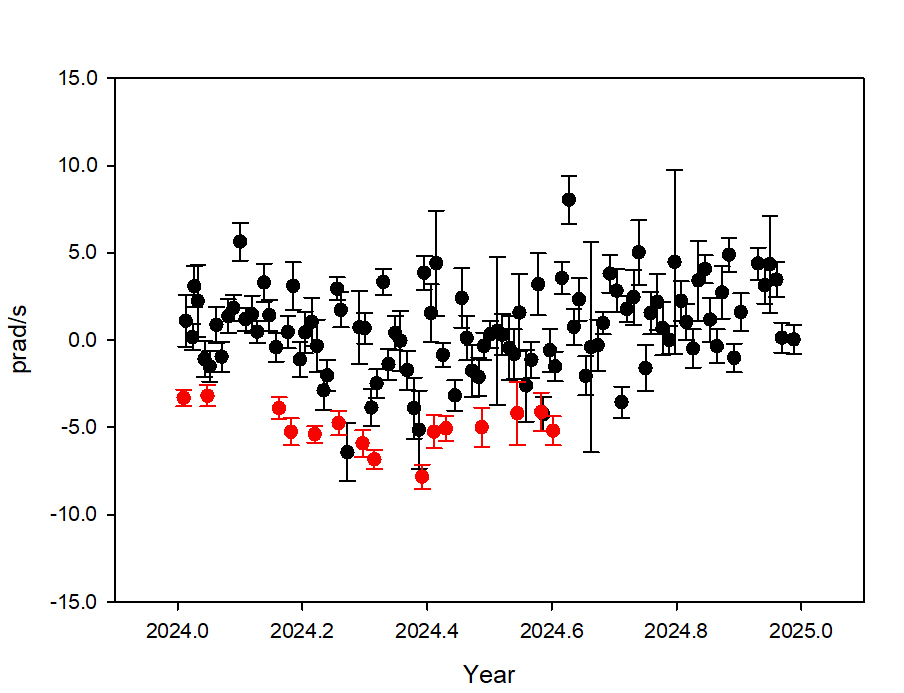}
    \caption{Daily corrections to the $\Omega$ nominal value from  IVS-R1 and IVS-R4 experiments (black dots) and VGOS experiments (red dots) during 2024.  }
    \label{fig12}
\end{figure}

Fig.~\ref{fig12} shows the estimates from all IVS-R1 and IVS-R4 experiments during 2024, compared with a set of VGOS experiments. The VGOS results display a smaller formal error and fewer range-to-range variations compared to the S/X counterparts. However, the VGOS estimates are biased with respect to the S/X estimates by about 5 prad/s. This suggests that the nominal fringe frequency $f$ for VGOS observations used in Eq.~(\ref{Eq1}) requires some correction. Once this frequency issue is addressed, VGOS data will provide a highly effective tool for monitoring the Earth's rotation angular velocity with record temporal resolution.

\section{Conclusion}

In conclusion, conventional astronomical methods do not allow the precise determination of the direction of Earth's axis of rotation, as there is no physical marker in the sky indicating this direction.  As a result, it is often assumed that it is fundamentally impossible to measure the direction of the axis using astronomical methods. However, geodetic VLBI, being a non-standard astronomical technique, is capable of measuring the motion of the 'rotating platform' (the Earth) using the Doppler effect, as presented in Eq.~(\ref{Eq1}). This capability, previously neglected, enables the kinematic determination of the direction of the Earth's axis of rotation using Eq.~(\ref{Eq2}). This approach defines the location of an imaginary marker on the sky, which can be used for practical applications. The accuracy is comparable to the accuracy of ring laser measurements, and, moreover, the temporal resolution of the observational time series is not limited to the standard 24-hour interval. Instead, the resolution can be reduced to 1-2 hours, provided a sufficient number of observations are available. A possible application of this new method is the monitoring of the Earth's rotational velocity in near real-time mode.

\begin{acknowledgements}
This paper is published with the permission of the CEO, Geoscience Australia. The authors acknowledge use of VLBI data obtained within the framework of the International VLBI Service for Geodesy and Astrometry (IVS). Length of Day (LOD) results published by the International Earth Rotation Service (IERS). 
I am grateful to Dr. Christian Bizouard from Paris Observatory and an anonymous reviewer for his comments and interesting discussion.

\end{acknowledgements}

\bibliographystyle{aa}
\bibliography{titov}
%\end{thebibliography}

\begin{appendix} 

\onecolumn

\section{Relativistic model for delay rate}

Relativistic model for the observed geodetic VLBI fringe frequency Eq.~(\ref{Eq1}) is calculated by differentiation of the group delay relativistic model from the IERS Conventions 2010 ~(\cite{iers10}, ~\cite{Kopeikin1990}, ~\cite{Klioner2003}, ~\cite{Soffel_2017}) so the final equation could be presented as follows

\begin{equation}
\begin{aligned}
\frac{\Delta f_{21}}{f} = -\frac{((w_{2} - w_{1}) \cdot s)}{c}\cdot
\left(1-\frac{2GM}{c^2r} - \frac{(V\cdot s)^2}{c^2} + \frac{V^2}{c^2} - \frac{(w_{2}\cdot s)}{c} -  \frac{(w_{2}\cdot V)}{c^2} \right)+
\frac{(b_{21}\cdot s)(a\cdot s)}{c^2}  + \frac{(b_{21}\cdot s)(a_{2}\cdot s)}{c^2}- \\
-\frac{(b_{21}\cdot a)}{c^2} 
- \frac{((w_{2} - w_{1}) \cdot V)}{c^2} + \frac{((w_{2} - w_{1}) \cdot V)(V\cdot s)}{2c^3} 	\label{sup_Eq1} % Use a logical label
\end{aligned}
\end{equation}
%\end{multline}

where $w_{1}$ and $w_{2}$ are the vectors of geocentric velocities of the radio telescopes, $r$ is the geocentric 
distance to the Solar centre, $s$ is the vector of direction to the radio source, $b_{21}$ is the baseline 
vector, $c$ is the speed of light, $V$ is the the vector of the Earth's barycentric velocity, $a$ is the vector of the Earth's barycentric 
acceleration, $a_{2}$ is the vector of the geocentric acceleration of the second station, $G$ is the 
gravitational constant, $M$ is the Solar mass, and $(\cdot)$ denotes the dot product.

The effect of general relativity is calculated in the analytical form based on the model of the 
light deflection angle (\cite{Titov_2015})

\begin{equation}
	\tau_{21} = \frac{2GM\left((b_{21}\cdot N)-(b_{21}\cdot s)(N\cdot s)\right)}{c^3(|r_{2}| - (r_{2}\cdot s))}.
	\label{Sup_Eq2} % Use a logical label
\end{equation}

Here $N$ is the unit vector towards the Sun's centre from the “second” telescope, i.e. $N = \frac{r_{2}}{|r_{2}|}$.

Several vectors are variable in Eq.~(\ref{Sup_Eq2}), but only the baseline vector changes rapidly enough to 
produce a measurable frequency shift for any radio source at any angular distance from the Sun. The corresponding model for the fringe frequency is developed by differentiation of the 
baseline vector in Eq.~(\ref{Sup_Eq2})

\begin{equation}
\begin{aligned}
	\frac{\Delta f_{21}}{f}=\frac{2GM((w_{2} - w_{1})(N-s(N\cdot s))}{c^3(|r_{2}| - (r_{2}\cdot s))} 
    = \frac{2GM(w_{2} - w_{1})[s\times [ s\times N]]}{c^3(|r_{2}| - (r_{2}\cdot s))} = \frac{2GM[\Omega \times b_{21}] \cdot [s\times [ s\times N]]}{c^3(|r_{2}| - (r_{2}\cdot s))}
	\label{sup_Eq3} % Use a logical label
\end{aligned}
\end{equation}
\\
\\

\section{Troposphere modeling for delay rate}

The wet troposphere modeling for the group delay is given by \citep{Boehm2006}

\begin{equation}
	\tau_{w}(z) = m_{w}(z)\tau_{w}(z_{0})+m_{w}(z)(G_{N}\cos A + G_{E}\sin A)
	\label{Sup_Eq4} % Use a logical label
\end{equation}

This model includes three parameters - zenith delay $\tau_{w}(z_{0})$ for the zenith direction ($z_{0} = 0$) and two gradients (North-South and East-West), $G_{N}$ and $G_{E}$, and the mapping function $m_{w}(z)$ to consider the actual zenith distance $z$ of the observed radio source.\\
The corresponding model for the delay rate includes also derivatives for the zenith delay rate and both gradients

\begin{equation}
\begin{aligned}  
\left(\frac{\Delta f}{f}\right)_{w}(z) = \dot m_{w}(z)\tau_{w}(z_{0})+
    m_{w}(z)\left(\frac{\Delta f}{f}\right)_{w}(z_{0})
    +  \dot m_{w}(z)\tan(z)(G_{N}\cos A + G_{E}\sin A) + \\
    + m_{w}(z)\tan(z)(\dot G_{N}\cos A + \dot G_{E}\sin A -
     G_{N}\dot A\sin A + G_{E}\dot A\cos A) 
      +  m_{w}(z)\frac{\dot z}{\cos^2 z}(G_{N}\cos A + G_{E}\sin A)
	\label{Sup_Eq5} % Use a logical label
\end{aligned}
\end{equation}

where six parameters are estimated in total in Eq.~(\ref{Sup_Eq5}), including three new derivatives: $\left(\frac{\Delta f}{f}\right)_{w}(z_{0})$, $\dot G_{N}$, $\dot G_{E}$. 

\clearpage   
\end{appendix}
\end{document}